\documentclass[conference]{IEEEtran}
\IEEEoverridecommandlockouts
\usepackage{cite}
\usepackage{float}
\usepackage{hyperref}
\usepackage{amsmath,amssymb,amsfonts}
\usepackage{balance}
\usepackage{algorithmic}
\usepackage{graphicx}
\usepackage{textcomp}
\usepackage{marvosym,graphicx,subfloat,subfig,tikz}
\usepackage{graphicx,fancyhdr} 
\usepackage[numbers]{natbib}
\usepackage{natbib}
\usepackage{rotating}
\usepackage[utf8]{inputenc} 
\usepackage[T1]{fontenc}    
\usepackage{url}            
\usepackage{booktabs}       
\usepackage{amsfonts}       
\usepackage{xcolor}
\def\BibTeX{{\rm B\kern-.05em{\sc i\kern-.025em b}\kern-.08em
    T\kern-.1667em\lower.7ex\hbox{E}\kern-.125emX}}
\begin{document}

\title{Predominant Musical Instrument Classification based on Spectral Features
}

\author{\IEEEauthorblockN{
		Karthikeya Racharla, Vineet Kumar\thanks{The work was done when authors were doing coursework at Indian Statistical Institute, Kolkata India.}, Chaudhari Bhushan Jayant, Ankit Khairkar and Paturu Harish
	}
\IEEEauthorblockA{{
		Indian Institute of Technology, Kharagpur India
		 \thanks{Corresponding Author Email: vineetkba2021@email.iimcal.ac.in}
	} \\
racharlakba2021@email.iimcal.ac.in, vineetkba2021@email.iimcal.ac.in, \\chaudharibba2021@email.iimcal.ac.in, khairkaraba2021@email.iimcal.ac.in, paturuhba2021@email.iimcal.ac.in	
}
}
\renewcommand{\headrulewidth}{0pt} 
\maketitle
\thispagestyle{fancy}

\begin{abstract}
This work aims to examine one of the cornerstone problems of Musical Instrument Retrieval (MIR), in particular, instrument classification. IRMAS (Instrument recognition in Musical Audio Signals) data set is chosen for this purpose. The data includes musical clips recorded from various sources in the last century, thus having a wide variety of audio quality. We have presented a very concise summary of past work in this domain. Having implemented various supervised learning algorithms for this classification task, SVM classifier has outperformed the other state-of-the-art models with an accuracy of 79\%. We also implemented Unsupervised techniques out of which Hierarchical Clustering has performed well. 
\end{abstract}

\begin{IEEEkeywords}
Musical Instrument Retrieval, Instrument Recognition, Spectral Analysis, Signal Processing
\end{IEEEkeywords}

\section{Introduction}
Music is to the soul what words are to the mind. With the advent of massive online streaming content, there is a need for on-demand music search that could be managed and stored easily. Also, audio tagging has become a challenge to explore. Music Information Retrieval (MIR) is the interdisciplinary research focused on retrieving information from music. Past the commercial implications, the development of robust MIR systems will contribute to a myriad of applications that include Recommender systems, Genre Identification and Catalogue Creation thus making the entire catalogue manageable and accessible with ease.

\subsection{Related Works} Instrument recognition is widely studied problem from various perspectives. \citet{essid2004} studied the classification of five different woodwind instruments. Mel frequency cepstral coefficient(MFCC) features were extracted from the training tracks as they were found helpful for classification based on tremolo, vibrato and sound attack. PCA was performed on the MFCC features for dimension reduction before feeding the transformed features to Gaussian Mixture model (GMM) and support vector machine (SVM) classification. GMM with 16, 32 Gaussian components were used, which resulted in better classification accuracy for the later. SVM was also performed with linear and polynomial kernels where the former was found to be efficient.

\citet{heittola2009musical} proposed a unique way to identify musical instrument from a polyphonic audio file. Training data consists of 19 different musical instruments. In the pre-processing of the audio file, multiple decomposition techniques are discussed such as Independent component analysis (ICA) and non-negative matrix factorization(NFM). The later provided a better separation of signals from the mixture of different sources of sound in the audio sample. Then Mel-frequency cepstral coefficient features were extracted from the reconstructed signals and fed to Gaussian Mixture model for classification.

\citet{han2017deep}'s approach to identify instrument revolved around extracting features from mel-spectrogram using convolutional layers of CNN. Mel spectrogram is the image containing information about playing style, frequency of sound excerpt and various spectral characteristics in music. The input given to the CNN is the magnitude of mel-frequency spectrogram which is compressed using natural logarithm. Various sampling techniques and transformations are performed to extract most of the information from sound excerpt .For more details one can refer \cite{han2017deep}. CNN architecture is proposed to identify instrument which comprised of convolutional layer which extracts feature from spectrogram automatically and max pooling is used for dimensionality reduction and classification. After experimenting with various activation functions, `ReLU' (alpha = 0.33) gave the best classification result with the overall F score of 0.602 on IRMAS training data.

\citet{hershey2017cnn} researched on Musical Instrument recognition in video clips. Their work primarily revolved around comparison of different neural network architectures based on accuracy. It has been observed that ResNet-50 yields the better result amongst Fully Connected, VGG, AlexNet architectures. A new data set Youtube-100M was created for this study.

\citet{toghiani2017musical} collected different music samples, transformed them into frequency domain and trained using the ANN model. Studies were carried out considering different circumstances – Complete music sample, using only the Attack, all other characteristic of music sample except Attack, the primary 100Hz frequency spectrum and the subsequent 900Hz of the same spectrum. The advantage of choosing frequency domain over time is to discretize the music sample directly, which is otherwise continuous. This would ease out the pre-processing effort.

According to \citet{Eronen2000}, Timbre, perceptually, is the \textit{colour} of a sound. Experiments have sought to construct a low-dimensional space to accommodate similarity ratings. Efforts are then made to interpret these ratings acoustically or perceptually.
The two principal dimensions here are spectral centroid and rise time. Spectral centroid corresponds to the perceived brightness of sound. Rise time measures the time difference between the start and the moment of highest amplitude.

\citet{deng2008study} have shown instruments usually have some unique properties that can be described by their harmonic spectra and their temporal and spectral envelopes. They have shown only the first few coefficients are enough for accurate classification.

\citet{murthy2018content} ascertained and critically reviewed the methods of extracting music related information given an audio sample. Emphasis was given on real data sets that are publicly available and gained popularity in the field of Music Information Retrieval. Areas covered under the study involve Music Similarity and Indexing, Genre, artist and raga identification along with music emotion classification. The research area finds its applications particular to personalized music cataloging and recommendations.
\bibliographystyle{plainnat}

\subsection{Our Contribution}
\begin{figure}[h!]
	\centering
	\subfloat[EDM ]{\label{fig:edm}\includegraphics[width=45mm]{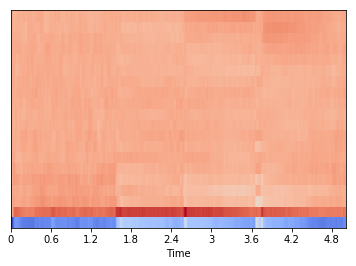}}
	\subfloat[Guitar (Accoustic) ]{\label{fig:gac}\includegraphics[width=45mm]{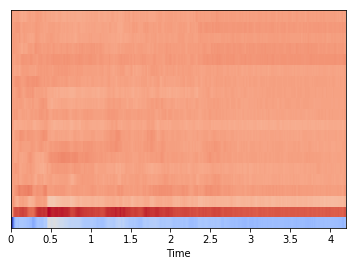}}
	\\
	\subfloat[Key Board ]{\label{fig:key}\includegraphics[width=45mm]{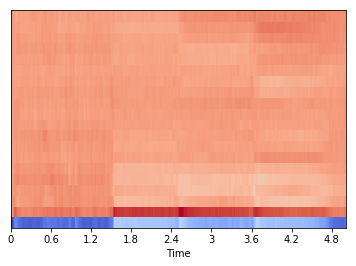}}
	\subfloat[Organ ]{\label{fig:org}\includegraphics[width=45mm]{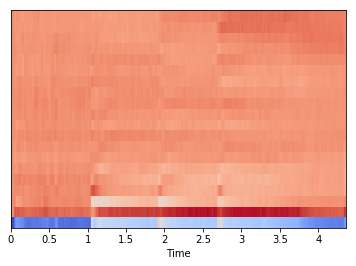}}
	\caption{Same note (audio) played by various instruments and their Spectograms}
	\label{fig:instr}
\end{figure}

To solve any classification problem of multimedia content, the most important thing is to figure out --- how to extract features from a given audio/video file. While dealing with our audio data set, we found that despite having the same notes of sound, the Spectrogram differs based on the instrument from which the note originates. As an illustration, we recorded the same note with four different instruments and generated the corresponding spectrograms, shown in the figure \ref{fig:instr}. It is evident that we can use this property of the spectrogram to predict the instrument used while playing a particular sound clip.

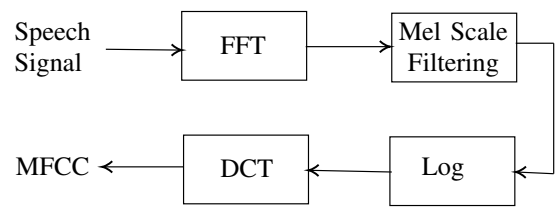
\begin{figure}[hb]
	\centering

\tikzset{every picture/.style={line width=0.45pt}} 
\begin{tikzpicture}[x=0.35pt,y=0.65pt,yscale=-1,xscale=1]

\draw   (200,11) -- (335,11) -- (335,51) -- (200,51) -- cycle ;
\draw   (427,12) -- (562,12) -- (562,52) -- (427,52) -- cycle ;
\draw   (425,84) -- (560,84) -- (560,124) -- (425,124) -- cycle ;
\draw   (202,83) -- (337,83) -- (337,123) -- (202,123) -- cycle ;
\draw    (118,33) -- (198,33.47) ;
\draw [shift={(200,33.48)}, rotate = 180.34] [color={rgb, 255:red, 0; green, 0; blue, 0 }  ][line width=0.75]    (10.93,-3.29) .. controls (6.95,-1.4) and (3.31,-0.3) .. (0,0) .. controls (3.31,0.3) and (6.95,1.4) .. (10.93,3.29)   ;

\draw    (334,31.48) -- (423,31.48) ;
\draw [shift={(425,31.48)}, rotate = 180] [color={rgb, 255:red, 0; green, 0; blue, 0 }  ][line width=0.75]    (10.93,-3.29) .. controls (6.95,-1.4) and (3.31,-0.3) .. (0,0) .. controls (3.31,0.3) and (6.95,1.4) .. (10.93,3.29)   ;

\draw    (427,104.48) -- (339,103.51) ;
\draw [shift={(337,103.48)}, rotate = 360.64] [color={rgb, 255:red, 0; green, 0; blue, 0 }  ][line width=0.75]    (10.93,-3.29) .. controls (6.95,-1.4) and (3.31,-0.3) .. (0,0) .. controls (3.31,0.3) and (6.95,1.4) .. (10.93,3.29)   ;

\draw    (202,101.48) -- (116,101.48) ;
\draw [shift={(114,101.48)}, rotate = 360] [color={rgb, 255:red, 0; green, 0; blue, 0 }  ][line width=0.75]    (10.93,-3.29) .. controls (6.95,-1.4) and (3.31,-0.3) .. (0,0) .. controls (3.31,0.3) and (6.95,1.4) .. (10.93,3.29)   ;

\draw    (561,29.48) -- (602,29.48) ;

\draw    (602,29.48) -- (602,105.48) ;

\draw    (602,105.48) -- (563,104.53) ;
\draw [shift={(561,104.48)}, rotate = 361.4] [color={rgb, 255:red, 0; green, 0; blue, 0 }  ][line width=0.75]    (10.93,-3.29) .. controls (6.95,-1.4) and (3.31,-0.3) .. (0,0) .. controls (3.31,0.3) and (6.95,1.4) .. (10.93,3.29)   ;

\draw (60.5,33) node  [align=left] {Speech \\Signal};
\draw (266.5,31) node  [align=left] {FFT};
\draw (493.5,33) node  [align=left] {Mel Scale \\ \ Filtering};
\draw (482.5,103) node  [align=left] {Log};
\draw (270.5,102) node  [align=left] {DCT};
\draw (60.5,101) node  [align=left] {MFCC};

\end{tikzpicture}
\caption{MFCC Calculation Scheme}
\label{mfc}
\end{figure}

For Spectral Analysis, MFCC \cite{logan2000mel} is the best choice. According to Wikipedia\cite{wiki:00} ``the mel-frequency cepstrum (MFC) is a representation of the short-term power spectrum of a sound, based on a linear cosine transform of a log power spectrum on a nonlinear mel-scale of frequency''. Mel is a number that links to a pitch, which is analogous to how a frequency is described by a pitch. The basic flow of calculating the MFC Coefficients is outlined in Fig: \ref{mfc}.
\\

\vspace{0.1in}

The mathematical formula for frequency-to-mel transform is given as: $$m = 2595\,\log_{10}\left(1+\frac f{700}\right).$$

MFCCs are obtained by transforming frequency (hertz) scale to mel scale. Typically, MFCC coefficients are numbered from the $0$th to 20th order and the first 13 coefficients are sufficient for our classification task. The lower order cepstral coefficients are primary representatives of the instrument. Though coefficients of the higher order give more fine tuned spectral details, choosing greater number of cepstral coefficients lands us in models of increased complexity. This would in turn require more training data for the estimation of model parameters.

In this work, we present a spectral feature based methodology for musical instrument classification. The rest of our paper is organised as follows --- Section: \ref{method} describes the dataset used, feature extraction approach and various classifier training techniques we used in our study. Section: \ref{result} presents the evaluation metric we have chosen, their interpretation and various performance wise study.  Finally Section: \ref{end} concludes the paper with few possible directions for future research.

\section{Methodology \& Our Approach}

\label{method}
\subsection{Data set}
IRMAS (Instrument recognition in Musical Audio Signals)\cite{bosch2012comparison} data has been used in our study. This data is polyphonic, hence using this dataset helps in building a robust classifier. The data consists of .wav files of 3 seconds duration of many instruments, eleven to be exact. We have chosen six of these instruments for recognition. Our data has 3846 samples of music running into about three hours, giving sufficient data for training and testing purposes as well. In addition, the data consists of multiple genres including country folk, classical, pop-rock and latin soul. Inclusion of these multiple genres could lead to better training. The data has been downloaded from \url{https://www.upf.edu/web/mtg/irmas}. Number of audio samples per instrument class is reproduced in table \ref{tab:ic}.

\begin{table}[h!]
	\centering
	
	\begin{tabular}{@{}lcc@{}}
		\textbf{Instrument} & \textbf{Number of Samples} & \multicolumn{1}{l}{\textbf{Clip Length (in sec)}}           \\\midrule
		Flute               & 451                        & 1,353                                                        \\
		Piano               & 721                        & 2,163                                                        \\
		Trumpet             & 577                        & 1,731                                                        \\
		Guitar              & 637                        & 1,911                                                        \\
		Voice               & 778                        & 2,334                                                        \\
		Organ               & 682                        & 2,046                                                        \\
		\textbf{Total}               & 3,846                       & 11,538 (3 hr 12 min) 
	\end{tabular}
	\vspace{0.1in}
	\caption{Instrument Samples and Clip Length}
	\label{tab:ic}
\end{table}

Code associated with our study is available for public use
\url{https://github.com/vntkumar8/musical-instrument-classification}.
\subsection{Feature Extraction}
\citet{deng2008study} have shown that for achieving more accurate classification of musical instruments, it is essential to extract more complicated features apart from MFCC. Hence, we considered other features like Zero-crossing rate, Spectral centroid, Spectral bandwidth and Spectral roll-off during our feature extraction using Librosa. 

\begin{itemize}
	\item \textbf{Zero Crossing Rate} (ZCR) indicates the rate at which the signal crosses zero.
	\item \textbf{Spectral Centroid} (SC) is a measure to indicate the center of mass of the spectrum being located, featuring the impression of brightness characteristic of a sample.  SC\cite{zhang2001audio}, is the ratio of the frequency weighted 	magnitude spectrum with unweighted magnitude spectrum.
	\item \textbf{Spectral bandwidth} (SB) gives the weighted average of the frequency signal by its spectrum.
	\item \textbf{Spectral roll-off} (SR) is the frequency under which a certain proportion of the overall spectral energy belongs to. Formally, SR is defined as the qth percentile
	of the power spectral distribution. 
\end{itemize}.  

To extract features from the audio files, we explored the standard libraries. We had two options -- Librosa\cite{mcfee2015librosa} and Essentia\cite{esen}. We used both of them in a Python-based Intel’s Jupyter platform and scikit-learn \cite{scikit} framework. Scikit-learn is an easy-to-use and open-source Machine Learning Library that supports most of the Supervised \& Unsupervised  Classification techniques.

Librosa is the Python package used for music and audio data analysis. Some important functions of librosa include Load, Display and Features. `Load' loads an audio file as floating point time series. `Display' provides visualizations such as waveform, spectrogram using matplotlib. `Features' is used for extraction and manipulation of MFCC and other spectral features. MFCCs are obtained by transforming from frequency (hertz) scale to mel-scale.

On the other hand, `Essentia' is an open-source C++ based distribution package available under Python wrapper environment for audio-based musical information retrieval. This library computes spectral energy associated with mel bands and their MFCCs of an audio sample. Windowing procedure is also implemented in Essentia. It analyzes the frequency content of an audio spectrum by creating a short sound segment of a few milli-seconds for a relatively longer signal. By default, we used Hann window\cite{esen}. It is a smoothing window typically characterized by good frequency resolution and reduced spectral leakage. The audio spectrum is analyzed by extracting MFCCs based on the default inputs of hopSize (hop length between frames) and frame size. The default parameters for sampling rate is 44.1 kHz, hopSize of 512 and frame size of 1024 in Essentia. The features thus extracted from manifold segments of a sample signal are aggregated with their mean. They are then used as the features for each sample labeled with their instrument class.

Librosa is one of the first Python libraries introduced to extract features from audio data. Librosa is also widely used and has a more established community support than Essentia. In our work, Librosa has provided better accuracy in out-of-sample validation. Hence we preferred Librosa as it led to greater accuracy.

\subsection{Classifier Training}

We extracted the first 13 MFCC features using Librosa/Essentia. For each audio clip, we obtained 259 $\times$ 13 matrix features. We took the mean of all the columns to get the condensed feature providing us with 1 $\times$ 13 feature vector, along with five other features as mentioned above. We labeled each vector with the instrument class using scikit-learn's `labelencoder' function.

\paragraph{Supervised Classification Techniques}

We implemented  80-20 shuffled split for training and testing sets along with cross validation to avoid over-fitting. Then we used different supervised classification techniques to identify the predominant musical instrument from the audio file. Initially we started with logistic regression and decision tree classifier. Classification trees are usually prone to over-fitting, so it did not perform well on the test data. We also used bagging and boosting techniques to train the MFCC and spectral features. We tried Random Forest to control the over-fitting. With some parameter tuning, it provided us with the better classification. We also tried XGBoost on the same set of features and after fine-tuning the parameters, Gradient Boosting classified the instruments with an accuracy of 0.7. 

We also used Support vector Machine(SVM) Classifier to fit the extracted features. It outperformed the traditional classification techniques. We used `radial basis function' kernel for this non-linear classification. 

The RBF kernel for two samples is represented as $${\displaystyle K(\mathbf {x} ,\mathbf {x'} )=\exp \left(-{\frac {\|\mathbf {x} -\mathbf {x'} \|^{2}}{2\sigma ^{2}}}\right)}$$
We also fine-tuned penalty parameter \textit{C} and kernel coefficient `gamma'. This improved the overall accuracy on test data. We cross validated the accuracy score of 79.41\% with 10 folds. In terms of accuracy, Bagging and Boosting models such as random forest and XGboost performed better than traditional models such as classification trees and logistic regression, which infact makes sense also. Finally, SVM turned out to be more accurate than other classifiers.

	\begin{figure}[h!]
	\centering
	\subfloat[Logistic Regression ]{\label{figur:1}\includegraphics[width=45mm]{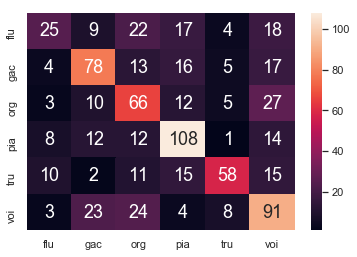}}
	\subfloat[Decision Tree ]{\label{figur:2}\includegraphics[width=45mm]{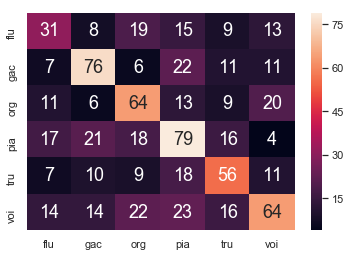}}
	\\\subfloat[Light GBM ]{\label{figur:3}\includegraphics[width=45mm]{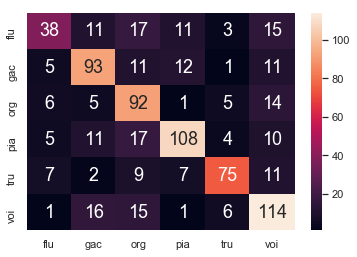}}
	\subfloat[XG Boost ]{\label{figur:4}\includegraphics[width=45mm]{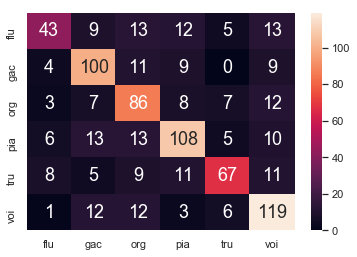}}
	\\\subfloat[Random Forest ]{\label{figur:5}\includegraphics[width=45mm]{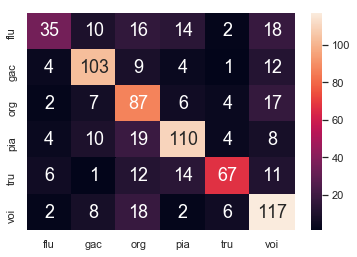}}
	\subfloat[SVM  ]{\label{figur:6}\includegraphics[width=45mm]{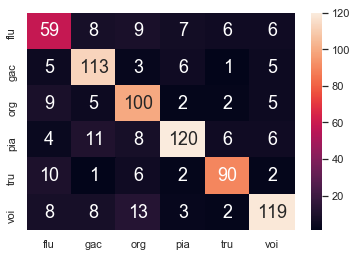}}
	\caption{Confusion Matrix for various supervised Algorithms}
	\label{figur}
\end{figure}

\paragraph{Unsupervised Classification Approach}

\begin{figure}[h!]
	\centering
	\includegraphics[width=0.85\linewidth]{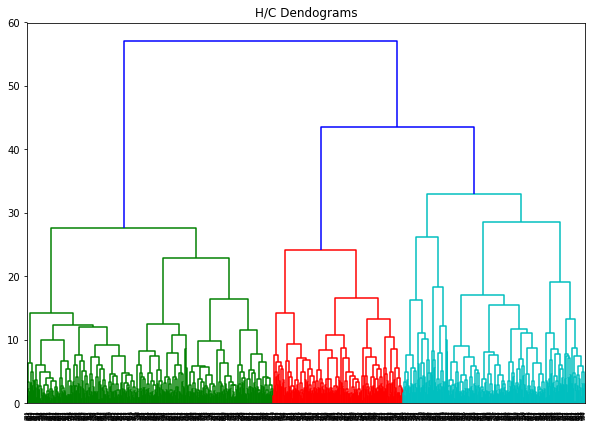}
	\caption{Hierarchical Cluster}
	\label{fig:hc}
\end{figure}

We tried unsupervised algorithms like `K-means' and `Hierarchical Clustering'. The Kmeans Clustering is not able assign appropriate cluster based on instruments. We took samples from 4 instruments and trained our K-means classifier. But the result was not at all acceptable. Perhaps the clusters were formed based on musical genres (viz. folk, country, blues etc.). However, Hierarchical Clustering perfomed reasonably well. It produced significant results when we cut the dendogram at the 30th level as shown in Fig: \ref{fig:hc}.


		\begin{table}[h!]					
			\centering\resizebox{9cm}{!} {
			\begin{tabular}{@{}l|lll|lll|lll@{}}
				\toprule
				& \multicolumn{3}{c}{\textbf{Logistic Regession}}                        & \multicolumn{3}{c}{\textbf{Decision Tree}}                             & \multicolumn{3}{c}{\textbf{LGBM}}                                      \\ \midrule
				Instrument & \multicolumn{1}{c}{P} & \multicolumn{1}{c}{R} & \multicolumn{1}{c}{F1} & \multicolumn{1}{c}{P} & \multicolumn{1}{c}{R} & \multicolumn{1}{c}{F1} & \multicolumn{1}{c}{P} & \multicolumn{1}{c}{R} & \multicolumn{1}{c}{F1}\\ 
				Flute      & 0.58                  & 0.39                  & 0.47                   & 0.43                  & 0.44                  & 0.43                   & 0.66                  & 0.59                  & 0.62                   \\
				Guitar      & 0.55                  & 0.59                  & 0.57                   & 0.53                  & 0.54                  & 0.53                   & 0.69                  & 0.73                  & 0.71                   \\
				Organ    & 0.44                  & 0.53                  & 0.48                   & 0.50                  & 0.46                  & 0.48                   & 0.59                  & 0.67                  & 0.63                   \\
				Piano     & 0.63                  & 0.57                  & 0.60                   & 0.60                  & 0.57                  & 0.58                   & 0.73                  & 0.68                  & 0.71                   \\
				Trumpet      & 0.58                  & 0.48                  & 0.52                   & 0.52                  & 0.50                  & 0.51                   & 0.72                  & 0.54                  & 0.62                   \\
				Voice      & 0.51                  & 0.61                  & 0.56                   & 0.50                  & 0.55                  & 0.52                   & 0.63                  & 0.74                  & 0.68                   \\ 
			\end{tabular}}
			
			\resizebox{9cm}{!} {
			\begin{tabular}{@{}l|lll|lll|lll@{}}
				\toprule
				& \multicolumn{3}{c}{\textbf{XG Boost}}                                  & \multicolumn{3}{c}{\textbf{RF}}                                        & \multicolumn{3}{c}{\textbf{SVM}}                                       \\ \midrule
				Instrument & \multicolumn{1}{c}{P} & \multicolumn{1}{c}{R} & \multicolumn{1}{c}{F1} & \multicolumn{1}{c}{P} & \multicolumn{1}{c}{R} & \multicolumn{1}{c}{F1} & \multicolumn{1}{c}{P} & \multicolumn{1}{c}{R} & \multicolumn{1}{c}{F1} \\
				Flute      & 0.66                  & 0.59                  & 0.62                   & 0.72                  & 0.48                  & 0.58                   & 0.63                  & 0.63                  & 0.63                   \\
				Guitar      & 0.72                  & 0.71                  & 0.71                   & 0.72                  & 0.75                  & 0.74                   & 0.79                  & 0.84                  & 0.81                   \\
				Organ    & 0.58                  & 0.69                  & 0.63                   & 0.61                  & 0.72                  & 0.66                   & 0.78                  & 0.77                  & 0.78                   \\
				Piano     & 0.71                  & 0.72                  & 0.71                   & 0.73                  & 0.72                  & 0.72                   & 0.77                  & 0.76                  & 0.77                   \\
				Trumpet      & 0.75                  & 0.53                  & 0.62                   & 0.74                  & 0.54                  & 0.62                   & 0.78                  & 0.67                  & 0.72                   \\
				Voice      & 0.65                  & 0.74                  & 0.69                   & 0.63                  & 0.80                  & 0.70                   & 0.79                  & 0.85                  & 0.82                   \\ \bottomrule
			\end{tabular}}
			\vspace{0.1 in }
			\caption{Precision, Recall \& F1 Score for  various Supervised Models}
			\label{tab:res}
		\end{table}

\begin{table}[h!]
	\centering
	
	\begin{tabular}{@{}lc@{}}
		\textbf{Model} & \multicolumn{1}{l}{\textbf{Accuracy}}           \\\midrule
		Logistic Regression               & 0.54                                                                              \\
		Decision Tree               & 0.52                        \\
		LGBM            & 0.66                       \\
		XG Boost              & 0.67                          \\
		Random Forest               & 0.68                         \\
		SVM              & 0.76                        \\
	\end{tabular}
	\vspace{0.1in}
	\caption{Accuracy of various Supervised Models as reported by Scikit Learn \cite{scikit}}
	\label{tab:avsm}
\end{table}


\section{Results \& Discussion}
\label{result}
\subsection{Evaluation Criteria}

The following evaluation metrics were used to judge the performance of the model

\begin{figure*}[ht]
	\centering
	
	\subfloat[Precision ]{\label{fig:precision}\includegraphics[width=90mm]{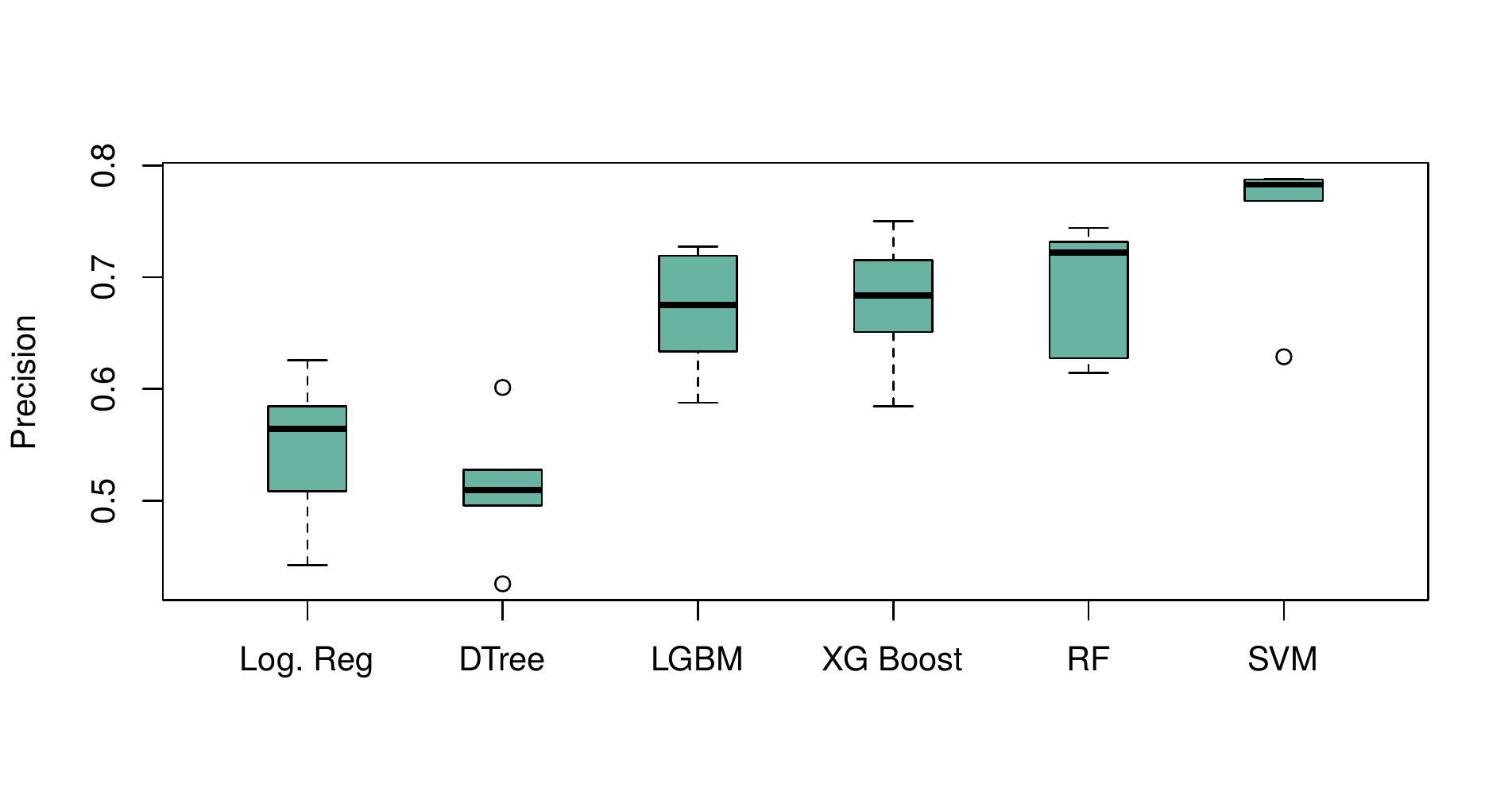}}
	\subfloat[Recall ]{\label{fig:recall}\includegraphics[width=90mm]{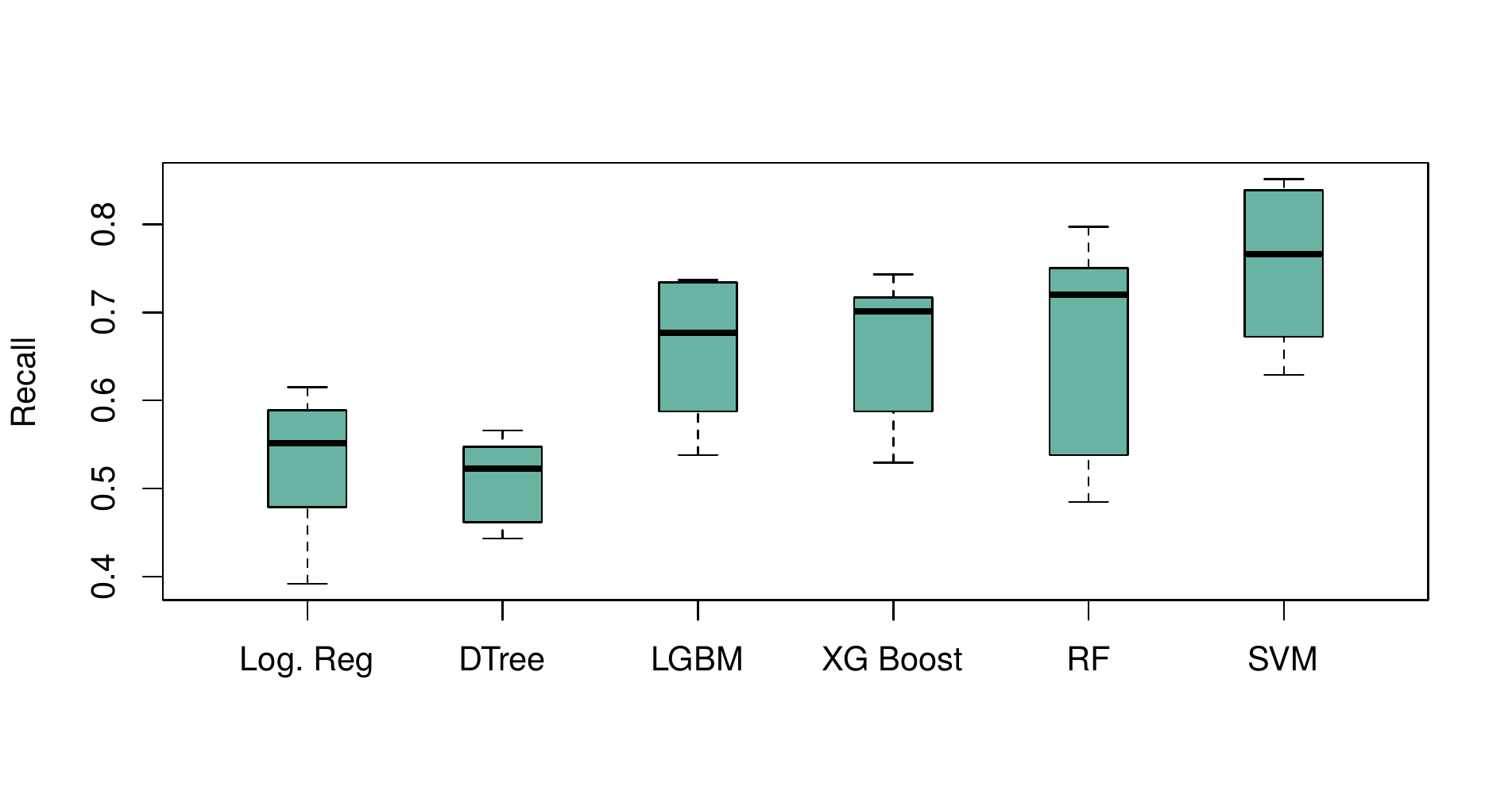}}
	\\
	\subfloat[F1 Score ]{\label{fig:f1}\includegraphics[width=90mm]{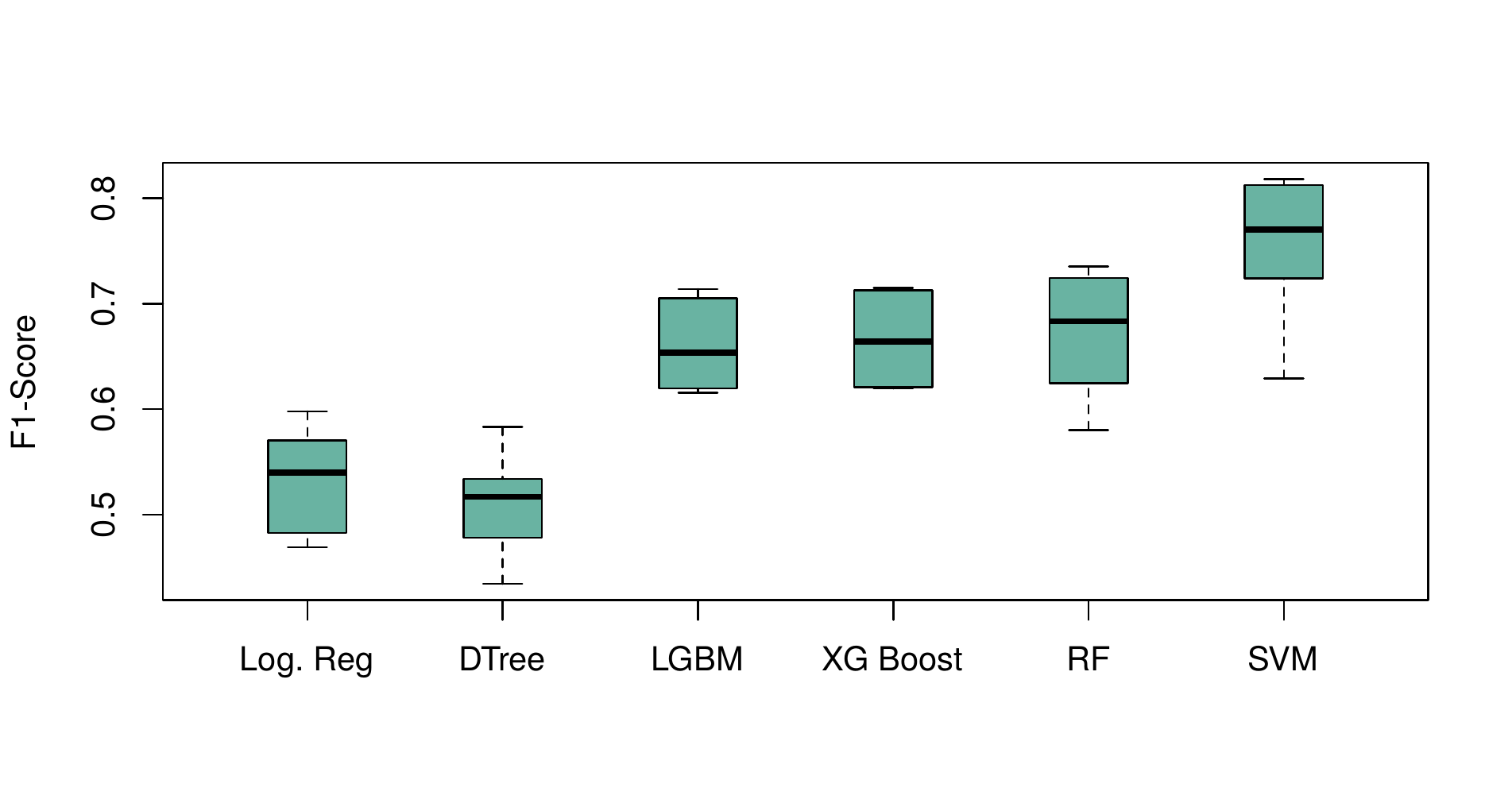}}
	\subfloat[Instrument wise classification ]{\label{fig:instrument-f1}\includegraphics[width=90mm]{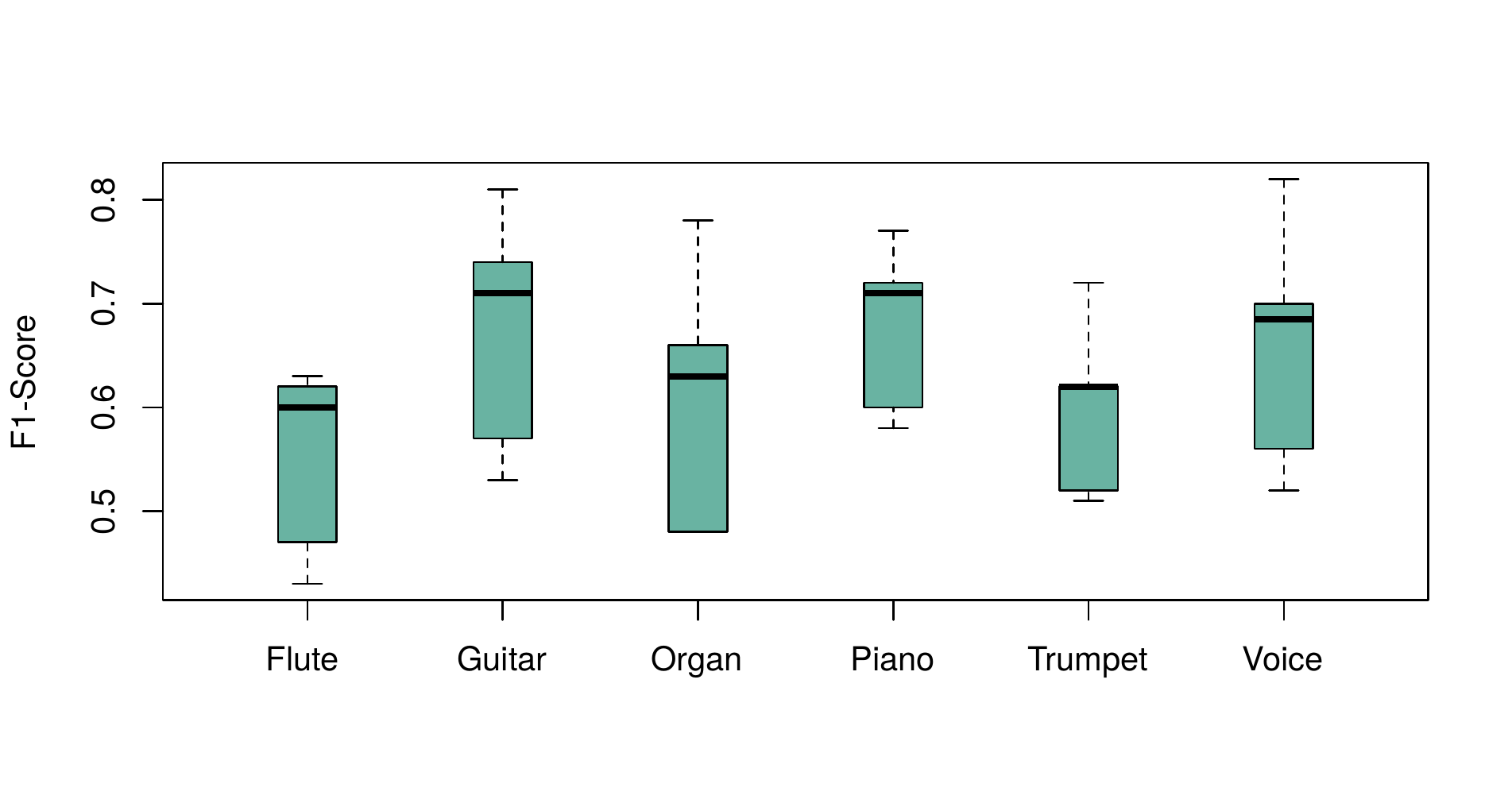}}
	\caption{Evaluation Metric for Various Supervised Algorithms}
\end{figure*}

\begin{itemize}
	\item \textbf{Precision} is the ratio $\frac{tp}{(tp + fp)}$, where $tp$ is the number of true positives and $fp$ the number of false positives. Precision intuitively describes the ability of the classifier not to label a false positive as positive. Precision for various models implemented is shown in box-plot see Fig: \ref{fig:precision}. 
	
	\item \textbf{Recall} is the ratio $\frac{tp}{(tp + fn)}$ where $tp$ is the number of true positives and $fn$ the number of false negatives. Recall is intuitively the ability of the classifier to identify all the positive samples. Fig: \ref{fig:recall} shows illustrative visualization of Recall for various supervised classifiers implemented. 
	
	\item \textbf{F1 score} can be interpreted as the harmonic mean of Precision and Recall. $$F1 = \frac{2 \times (\rm{precision} \times \rm{recall})}{ (\rm{precision} + \rm{recall})}$$
	
	\item We reported \textbf{Accuracy} as well. It disregard the class breakdown, and simply label each observation as either correct or incorrect classification. The accuracy is the proportion of correctly classified examples. It has been reported in Table: \ref{tab:avsm}
	\item \textbf{Confusion Matrix} evaluates the performance of a supervised classifier using a cross-tabulation of actual and predicted classes. The comparison for various models is shown in Fig: \ref{figur}
\end{itemize}

From the table \ref{tab:res}, it is clear that SVM yields the highest accuracy. However, it has been observed that the classifier is unable to distinguish between flute and organ properly. See Fig: \ref{fig:instrument-f1} the lower whisker of boxplot is evident of the fact that classifier is confusing. Samething is clear from Fig: \ref{figur} confusion matrix plot. The reason can be traced back to both being essentially the wind instruments. Their spectral envelopes are mostly identical in nature hence accurate classification is bit tricky. 

Fig: \ref{fig:precision} is the box plot of various supervised models. It presents the precision of models with respect to  various instruments. The six instruments and their precisions have been used to generate the box plot. In exacly same way for each of the instrument same thing has been done. Interpretation is similar for other Fig \ref{fig:recall} and \ref{fig:f1} which explains Recall and F1 measure respectively.

Fig: \ref{fig:instrument-f1} presents a different picture, it is box plot of various instruments. This means F-scores for various models were used in calculating the box representation of each instrument. 

\section{Future Scope}\label{end}
There is scope to use the same approach on a different data set. One can explore the idea of classifying Indian instruments like shehnai, jaltarang, dholak etc. Fundamental bottleneck for no concrete study in this direction might be non availability of labelled dataset. More libraries for extraction of MFCC features can be explored, as we implemented only two libraries viz. Librosa and Essentia. One may look at deep neural networks based approach. More features in addition to the MFCCs can be studied and extracted using signal processing techniques to improve the accuracy of instrument classification.

\section*{Acknowledgment}
The authors would like to thank the anonymous reviewers of SPIN 2020 for their constructive criticism and suggestions, which helped in substantially improving the technical and editorial quality of the paper. Authors would also like to thank Intel Devcloud for providing us a virtual machine
which we used in our experiments.

	\bibliographystyle{ieeetr}. \balance \bibliography{bibo}

\begin{thebibliography}{15}
\providecommand{\natexlab}[1]{#1}
\providecommand{\url}[1]{\texttt{#1}}
\expandafter\ifx\csname urlstyle\endcsname\relax
  \providecommand{\doi}[1]{doi: #1}\else
  \providecommand{\doi}{doi: \begingroup \urlstyle{rm}\Url}\fi

\bibitem[Bosch et~al.(2012)Bosch, Janer, Fuhrmann, and
  Herrera]{bosch2012comparison}
Juan~J Bosch, Jordi Janer, Ferdinand Fuhrmann, and Perfecto Herrera.
\newblock A comparison of sound segregation techniques for predominant
  instrument recognition in musical audio signals.
\newblock In \emph{ISMIR}, pages 559--564, 2012.

\bibitem[Deng et~al.(2008)Deng, Simmermacher, and Cranefield]{deng2008study}
Jeremiah~D Deng, Christian Simmermacher, and Stephen Cranefield.
\newblock A study on feature analysis for musical instrument classification.
\newblock \emph{IEEE Transactions on Systems, Man, and Cybernetics, Part B
  (Cybernetics)}, 38\penalty0 (2):\penalty0 429--438, 2008.

\bibitem[Eronen and Klapuri(2000)]{Eronen2000}
A.~Eronen and A.~Klapuri.
\newblock Musical instrument recognition using cepstral coefficients and
  temporal features.
\newblock In \emph{Proceedings of the Acoustics, Speech, and Signal Processing,
  2000. On IEEE International Conference - Volume 02}, ICASSP '00, pages
  II753--II756, Washington, DC, USA, 2000. IEEE Computer Society.
\newblock ISBN 0-7803-6293-4.
\newblock \doi{10.1109/ICASSP.2000.859069}.
\newblock URL \url{http://dx.doi.org/10.1109/ICASSP.2000.859069}.

\bibitem[Essid et~al.(2004)Essid, Richard, and David]{essid2004}
Slim Essid, Ga{\"e}l Richard, and Bertrand David.
\newblock Musical instrument recognition on solo performances.
\newblock In \emph{2004 12th European signal processing conference}, pages
  1289--1292. IEEE, 2004.

\bibitem[Han et~al.(2017)Han, Kim, Lee, Han, Kim, and Lee]{han2017deep}
Yoonchang Han, Jaehun Kim, Kyogu Lee, Yoonchang Han, Jaehun Kim, and Kyogu Lee.
\newblock Deep convolutional neural networks for predominant instrument
  recognition in polyphonic music.
\newblock \emph{IEEE/ACM Transactions on Audio, Speech and Language Processing
  (TASLP)}, 25\penalty0 (1):\penalty0 208--221, 2017.

\bibitem[Heittola et~al.(2009)Heittola, Klapuri, and
  Virtanen]{heittola2009musical}
Toni Heittola, Anssi Klapuri, and Tuomas Virtanen.
\newblock Musical instrument recognition in polyphonic audio using
  source-filter model for sound separation.
\newblock In \emph{ISMIR}, pages 327--332, 2009.

\bibitem[Hershey et~al.(2017)Hershey, Chaudhuri, Ellis, Gemmeke, Jansen, Moore,
  Plakal, Platt, Saurous, Seybold, et~al.]{hershey2017cnn}
Shawn Hershey, Sourish Chaudhuri, Daniel~PW Ellis, Jort~F Gemmeke, Aren Jansen,
  R~Channing Moore, Manoj Plakal, Devin Platt, Rif~A Saurous, Bryan Seybold,
  et~al.
\newblock Cnn architectures for large-scale audio classification.
\newblock In \emph{2017 ieee international conference on acoustics, speech and
  signal processing (icassp)}, pages 131--135. IEEE, 2017.

\bibitem[Logan et~al.()]{logan2000mel}
Beth Logan et~al.
\newblock Mel frequency cepstral coefficients for music modeling.

\bibitem[McFee et~al.(2015)McFee, Raffel, Liang, Ellis, McVicar, Battenberg,
  and Nieto]{mcfee2015librosa}
Brian McFee, Colin Raffel, Dawen Liang, Daniel~PW Ellis, Matt McVicar, Eric
  Battenberg, and Oriol Nieto.
\newblock librosa: Audio and music signal analysis in python.
\newblock In \emph{Proceedings of the 14th python in science conference},
  volume~8, 2015.

\bibitem[{MTG upf}(2019)]{esen}
{MTG upf}.
\newblock Essentia open-source library and tools for audio and music analysis,
  description and synthesis, 2019.
\newblock URL \url{https://essentia.upf.edu/documentation/index.html}.
\newblock [Online; accessed 23-November-2019].

\bibitem[Murthy and Koolagudi(2018)]{murthy2018content}
YV~Murthy and Shashidhar~G Koolagudi.
\newblock Content-based music information retrieval (cb-mir) and its
  applications toward the music industry: A review.
\newblock \emph{ACM Computing Surveys (CSUR)}, 51\penalty0 (3):\penalty0 45,
  2018.

\bibitem[Pedregosa et~al.(2011)Pedregosa, Varoquaux, Gramfort, Michel, Thirion,
  Grisel, Blondel, Prettenhofer, Weiss, Dubourg, Vanderplas, Passos,
  Cournapeau, Brucher, Perrot, and Duchesnay]{scikit}
F.~Pedregosa, G.~Varoquaux, A.~Gramfort, V.~Michel, B.~Thirion, O.~Grisel,
  M.~Blondel, P.~Prettenhofer, R.~Weiss, V.~Dubourg, J.~Vanderplas, A.~Passos,
  D.~Cournapeau, M.~Brucher, M.~Perrot, and E.~Duchesnay.
\newblock Scikit-learn: Machine learning in {P}ython.
\newblock \emph{Journal of Machine Learning Research}, 12:\penalty0 2825--2830,
  2011.

\bibitem[Toghiani-Rizi and Windmark(2017)]{toghiani2017musical}
Babak Toghiani-Rizi and Marcus Windmark.
\newblock Musical instrument recognition using their distinctive
  characteristics in artificial neural networks.
\newblock \emph{arXiv preprint arXiv:1705.04971}, 2017.

\bibitem[{Wikipedia contributors}(2019)]{wiki:00}
{Wikipedia contributors}.
\newblock Mel-frequency cepstrum --- {Wikipedia}{,} the free encyclopedia,
  2019.
\newblock URL
  \url{https://en.wikipedia.org/w/index.php?title=Mel-frequency_cepstrum&oldid=917928298}.
\newblock [Online; accessed 23-November-2019].

\bibitem[Zhang and Kuo(2001)]{zhang2001audio}
Tong Zhang and C-C~Jay Kuo.
\newblock Audio content analysis for online audiovisual data segmentation and
  classification.
\newblock \emph{IEEE Transactions on speech and audio processing}, 9\penalty0
  (4):\penalty0 441--457, 2001.

\end{thebibliography}
	
\end{document}